\documentclass[journal]{IEEEtran}
\IEEEoverridecommandlockouts
\usepackage[dvipsnames]{xcolor}
\usepackage{amsmath,amsthm,amssymb,amsfonts}
\usepackage{boxedminipage}
\usepackage{pgfplotstable}
\usepackage{float}
\usepackage{url}
\usepackage{eurosym}
\usepackage{cite}
\usepackage{algorithm}
\usepackage{algorithmic}

\usepackage{graphicx}
\usepackage{textcomp}
\usepackage{eurosym}

\usepackage[utf8x]{inputenc}
\usepackage{lmodern,textcomp}
\usepackage{epstopdf}

\DeclareMathOperator*{\argmin}{arg\,min}

\makeatletter
\def\ps@IEEEtitlepagestyle{%
  \def\@oddfoot{\mycopyrightnotice}%
  \def\@evenfoot{}%
}
\def\mycopyrightnotice{%
  {\footnotesize  
	\begin{minipage}{\textwidth}
	\centering
~\copyright~2021 IEEE. Personal use of this material is permitted.  
Permission from IEEE must be obtained for all other uses, in any current or future media, including reprinting/republishing this material for advertising or promotional purposes, creating new collective works, for resale or redistribution to servers or lists, or reuse of any copyrighted component of this work in other works.\hfill
	\end{minipage}
}
  \gdef\mycopyrightnotice{}
}

\def\BibTeX{{\rm B\kern-.05em{\sc i\kern-.025em b}\kern-.08em
    T\kern-.1667em\lower.7ex\hbox{E}\kern-.125emX}}
\begin{document}

\title{Decentralized Assignment of Electric Vehicles at Charging Stations Based on Personalized \textcolor{black}{Cost} Functions and Distributed Ledger Technologies}
\author{{Michela Moschella}\thanks{Michela Moschella and Emanuele Crisostomi are with the Department of Energy, Systems, Territory and Constructions Engineering, University of Pisa, Pisa, Italy (e-mail: michela.moschella@ing.unipi.it, emanuele.crisostomi@unipi.it)}, \IEEEmembership{Student Member, IEEE}, Pietro Ferraro\thanks{Pietro Ferraro and Robert Shorten are with the Dyson School of Engineering Design, Imperial College London, London, UK (e-mail: p.ferraro@imperial.ac.uk, r.shorten@imperial.ac.uk)},
{Emanuele Crisostomi, \IEEEmembership{Senior Member, IEEE}, and Robert Shorten},
\IEEEmembership{Senior Member, IEEE}}

\maketitle

\begin{abstract}
In this paper we propose a stochastic decentralized algorithm to recommend the most convenient Charging Station (CS) to Plug-in Electric Vehicles (PEVs) that need charging. 
In particular, we use different \textcolor{black}{cost} functions to describe the possibly different priorities of PEV drivers, 
such as the preference \textcolor{black}{to minimize charging costs, charging times, or the distance between them and the CS}. 
For this purpose, we leverage on an IoT architecture based on a permissioned Distributed Ledger Technology (DLT) to enforce compliance of drivers and reduces the occurrence of detrimental misbehaviours of drivers.  Extensive simulations performed with the mobility simulator SUMO in realistic city-wide networks have been provided to illustrate how the proposed PEV assignment procedure works in practice, and to validate its performance.
\end{abstract}


\section{Introduction}
\label{sec:introduction}

\subsection{Motivations}
\textcolor{black}{
The Internet of Things (IoT) has become, in recent years, a topic of great interest in academia and industry alike, thanks to the advent of smart homes, smart cities, and smart everything \cite{Khan}.
The interaction of numerous small and lightweight computing elements is expected to enable a number of distributed services and opportunities for innovation at various levels. 
From a daily life perspective, IoT is expected to affect a vast array of domains: housing \cite{Miori}, health-related applications \cite{Bhatt} and enhanced learning \cite{Farhan} are just a few examples of the areas that are going to be dramatically altered.
At the same time, at a higher level, IoT is expected to bring advantages to the community as a whole, by providing services such as smart mobility \cite{oldpaper}, detecting weather conditions \cite{Li} and monitoring surgery in hospitals \cite{Yeole}.} 

\textcolor{black}{
Concurrently with this ever-expanding interest in the IoT world, in the mobility sector it is possible to notice an increasing interest in electric vehicles (EVs), whose number of sales worldwide is rapidly increasing every year. While the original main driver towards the adoption of EVs was the decreasing of harmful tailpipe emissions in city centers, there is now a growing interest of power Distributed System Operators (DSOs) in utilizing EVs as mobile storage units, in the case of Plug-in EVs (PEVs), to improve the robustness and the resilience of the power grid. 
In addition, PEVs may be successfully used to facilitate the penetration of Renewable Energy (RE), to both mitigate their fluctuating energy generation, and fully exploit energy generated at peak times. 
It is thus clear that a smart management of the charging process of PEVs is required to fully exploit their potential, while avoiding possible inefficiencies for the distribution grid, e.g., in terms of voltage limits of the buses or thermal inefficiencies of electrical feeders and substations.
}

In this context, the connectivity between PEVs fleets, the related interactive infrastructures (e.g., charging stations (CSs)) and effective management of their interactions is an emerging application of the IoT technology. The intersection between these two emerging domains is thus the focus of this work. In particular, in this manuscript we investigate the problem of optimally assigning PEVs to fast CSs \textcolor{black}{(the \emph{PEV-CS assignment problem} from now on)}, taking into account personalized \textcolor{black}{cost} functions; actually, different individuals may be interested in choosing their preferred CS according to different reasons, which could range from minimum charging price, to minimum travel distance or minimum charging time. For this purpose, we shall leverage on IoT based technology to conveniently share the relevant information, solve the assignment problem in a decentralized manner, and further consider the compliance of users to actually travel towards the recommended CSs. 

{\color{black} To enunciate clearly the specific contributions of this paper, we note that our starting point is the companion paper \cite{bibref:hausler}. This paper was one of the first works to formulate the problem of decentralized charge-point assignment, and to suggest a solution. In this present paper we extend significantly the results of \cite{bibref:hausler}, by relaxing four basic assumptions that underpin work described therein.\newline

\begin{itemize}

\item[1.] {\bf Cooperation:} A strong assumption in \cite{bibref:hausler} is that CS owners cooperate with each other to achieve balancing. This assumption is, of course, very unrealistic. In the present setting, competitive competition between owners is considered.\newline

\item[2.] {\bf Homogeneity:} Another assumption in \cite{bibref:hausler} is that all PEV owners have the same preferences. Here, we consider that in reality some drivers may have different priorities, and in particular may be interested in minimizing charging times, or charging costs (that in fact, had not been considered in \cite{bibref:hausler}).\newline

\item[3.] {\bf Compliance:} The third assumption in \cite{bibref:hausler} is that car-owners comply ridgidly with assignments. This assumption, which is a cornerstone of many works in Smart Cities, is clearly 
not realistic as the issues around compliance in the context if Covid-19 clearly demonstrate. Distributed compliance strategies are not only clearly necessary in society, but also actually represent a challenging and important frontier topic in control theory, IoT, and other disciplines, requiring the design of compliance strategies that are fair to all participants, bias free, and preserve privacy. To the best of our knowledge, the novel DLT-based compliance method here is introduced for the first time in the context of CS assignment. We emphasize  that the issue of compliance in the context of this (and similar) problems is usually dealt with in an indirect manner using a {\em game theoretic} approach. In the game theoretic approach one seeks to create an equilibrium that incentives good behaviour. Our approach is fundamentally different. Here, we use a DLT architecture to create a personalised economic commitment mechanism using tokens to enforce compliance. It is worth also noting that this approach is extremely well grounded in behavioural economics by the theory of hyperbolic discounting \cite{HD}, but becomes extremely implementable using DLT's that do not require transaction fees. The DLT can also be used to enable other services such as secure ownership of exchange of data (which is not discussed in the present manuscript), and to ensure fairness and bias free operation.\newline

 \item[4] {\bf Risk:} The fourth assumption, based on Item 3. above, concerns the risk of non-compliance, its pricing, and its management. As we mentioned, the issue of compliance is ignored in \cite{bibref:hausler}  and indeed most previous works on this topic. A further contribution in this context is to tackle the compliance problem, and specifically how to price compliance, using a control theoretic approach; specifically, by wrapping a feedback loop around the compliance problem. By doing this the risk of a reduction in the quality of service can be controlled and managed in real time. Notice that solving the problem of economic commitment in this manner gives rise to many new mathematical challenges akin to stochastic economic string stability as enunciated in \cite{oldpaper}. These problems are however beyond the scope of the present paper which is developed for an IoT setting; the interested reader may however refer to \cite{oldpaper} for a preliminary formulation of some of these problems. \newline
 \end{itemize}

A further contribution beyond \cite{bibref:hausler}  is that the assignment problem is solved in a fully decentralized fashion.  Thus, neither PEVs have to exchange information with other PEVs, nor CSs require to exchange information with other CSs. This is convenient as CSs may be reluctant to reveal possibly sensitive information (e.g., the length of the queue of charging PEVs or charging prices), and the same is true for PEV owners (e.g., how much energy they need). These are additional cotributions over \cite{bibref:hausler}. Finally, it is also worth noting that this paper presents a unique perspective on the problem of charge point assignment, bringing ideas from different domains such as DLTs, decentralised optimization, control theory and unifies them in a cyber-physical framework to manage the increasing amount of electric vehicles in our cities and properly exploit their full potential.

\subsection{Organization}
The remainder of this paper is organized as follows. Section \ref{Sec: State of the art} revises the current state of the art.  Section \ref{Assignment_Algorithm} describes the PEV-CS assignment problem, and formulates the proposed decentralized optimization problem to solve it. Then, Section \ref{sec:DLTToken} presents the system architecture and the compliance analysis. The following Section \ref{sec:results} shows and discusses the obtained simulation results. Finally, Section \ref{conclusion} summarizes our main findings and outlines possible future extensions of our work.}

\section{State of the Art}
\label{Sec: State of the art}
It is known that uncontrolled charging of PEVs may lead to grid problems in terms of power losses and voltage deviations, especially in presence of a large penetration of PEVs in the vehicular fleet \cite{bibref:clement}. Accordingly, the impact of PEVs on the power grid has been widely investigated in a number of papers, see for instance \cite{bibref:rigas} where a comprehensive survey of artificial-intelligence based solutions is provided, and \cite{bibref:zheng} for deterministic methodologies like Model Predictive Control (MPC). In this context, the availability of measured data is extremely valuable to realistically model actual real drivers' charging patterns (e.g., \cite{bibref:flam}, \cite{bibref:zoep}) and design well-coordinated charging strategies \cite{bibref:kont}. Some works have rather focused on the minimization of the impact on the power grid \cite{bibref:ma}, while other works prioritized the waiting time at CSs \cite{bibref:hausler} or the minimization of polluting emissions and charging costs \cite{bibref:kont}.\\
\newline
Specifically, in this paper we are interested in the \emph{PEV-CS assignment problem}, which consists in recommending the most convenient CS to any vehicle requesting to be charged. Usually this problem is formulated as an optimization problem, where the best assignment is the one that minimizes a \textcolor{black}{cost} function of interest. For instance, in \cite{bibref:cheki} the optimal scheduling aimed at minimizing the waiting time of PEVs at each CS; in \cite{bibref:jin} a Lyapunov optimization approach was used to improve the utilization of RE and reduce charging costs. A combined cost function was used in \cite{bibref:etesami} to take into account traffic congestion, waiting time at CSs, battery constraints and also the energy price.\\
\newline
However, most of the aforementioned approaches (e.g., \cite{bibref:ma}, \cite{bibref:etesami} and \cite{bibref:zou})  may arguably have a limited effectiveness in practice. For instance, optimal solutions are static: as a Nash Equilibrium represents the optimal solution of a static system, a new element (that could be a CS or a PEV) entering or leaving the system may in general require to recompute the optimal solution. Also, such methodologies implicitly assume that the actors (here the PEVs) will follow the received optimal recommendations, as declining to follow the received recommendation has an impact on the optimality of the global solution. In addition, another important aspect that is often neglected in most of the literature is the behavior of the drivers, as not all drivers have the same priorities, and may take different decisions (e.g., the choice of a CS where charging) based on their own perception and interests. For instance, this aspect is investigated in \cite{bibref:etesami} and \cite{bibref:hu}, where prospect theory \cite{bibref:kahn} and cumulative prospect theory respectively are used to model real-life human choices, and in \cite{bibref:mehdi}, where the interaction of drivers with utilities and retailers is modeled as a function of their social classes (e.g., to predict the reaction of PEVs to discount charging fees).\\
\newline
Another aspect \textcolor{black}{to be taken} in consideration is the underlying architecture of the model; 
the IoT paradigm offers the ability to ensure security and efficiency of a system, with low communication and computation costs.
Nowadays, many researchers are applying the IoT communication to the electric vehicles world, with many specific applications; 
among various examples, \cite{bibref:zeng} proposes an E-AUA protocol for driverless electric cars, for a secure and efficient system while \cite{bibref:tang} exploits the IoT technology to jointly optimize the PEVs routing and the charging scheduling, by means of a distributed algorithm. 
In another example \cite{bibref:savari}, the authors focus on the PEVs charging problem, and propose a real-time solution to improve the charging scheduling of PEVs, without the intervention of a third-party and protecting the PEVs' user privacy. 
Unfortunately, the latter approach is centralized, because it is based on a database which contains CSs and PEVs data to be processed, while decentralized algorithm, generally harder to implement, are preferred for a more efficient and scalable system. 
In the context of this push towards decentralization, concurrent with this growth of interest in IoT, Distributed Ledger Technologies (DLTs) have been proposed as a solution for distributed consensus in a database \cite{Dorri}, in a peer to peer (P2P) network. 
In the aftermath of the 2008 financial crisis, Satoshi Nakamoto proposed in their whitepaper \cite{Nakamoto} the use of an architecture, called the Blockchain, to create a distributed digital currency, the Bitcoin. 
Since then, academia and industry have been studying this new technology to expand its scope of application beyond financial transactions. 
To name a few examples, DLTs have been proposed as an enabling technology for managing, controlling and securing interactions in a range of cyberphysical systems \cite{Samaniego}\cite{Novo}\cite{Panarello}. 
Due to their distributed and trustless nature, DLTs have characteristics that prove advantageous as a data transfer and transaction settlement system for the IoT domain \cite{IoTPaper}. 
Still, despite these advantages, Blockchain presents many limitations: the large energy cost of mining, long transaction approval times, transaction fees, and the inherent preference of miners to prioritize some transactions (generally ones associated with larger economic values)\cite{oldpaper}. 
These characteristics create a bottleneck when thousands of devices communicate with each other many times per second. 
While several Blockchain-based solutions have been proposed to overcome these problems \cite{Dorri2} \cite{Novo} \cite{Dorri}, in this paper we focus on a different kind of DLT architecture based on Directed Acyclic Graphs (DAGs) and more specifically on the one proposed in \cite{SPtoken} and \cite{TyrePaper}. 
This architecture seems to possess properties that make it suitable for the IoT domain; for a more thorough comparison between Blockchain-based DLTs and DAG-based DLTs, in the IoT context, the interested reader can refer to \cite{oldpaper}.

\begin{figure}[h!]
\centering
\includegraphics[width=\columnwidth]{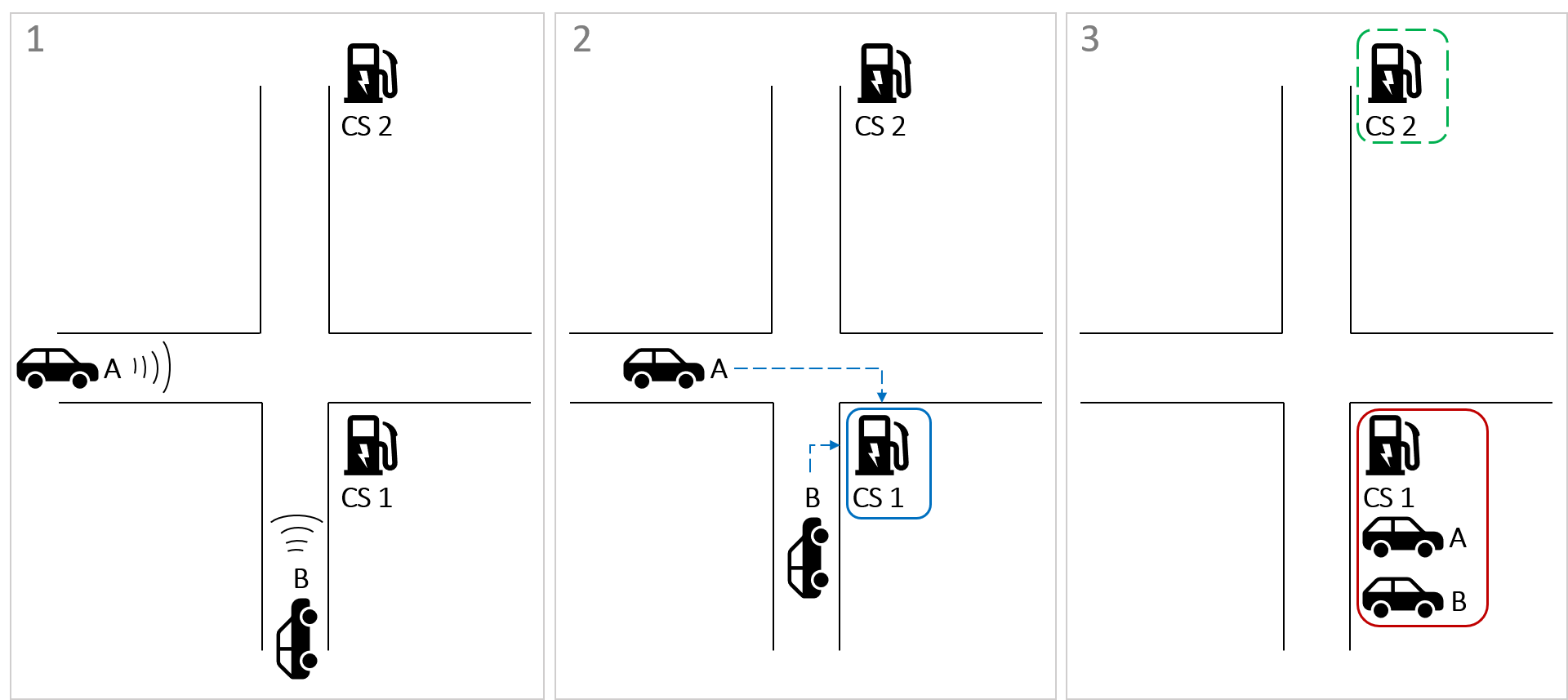}
\caption{\textcolor{black}{Suboptimality of CSs network usage: (left panel) cars A and B look for a CS; (central panel) the nearest station is CS 1 for both car A and car B; (right panel) given the relatively long time required for charging a PEV, the obtained solution with a queue of two vehicles at the same CS is less convenient than recommending a different CS to the two PEVs.}}
\label{fig:context}
\end{figure}

\textcolor{black}{\section{The PEV-CS Assignment Algorithm}
\label{Assignment_Algorithm}
We consider an urban network, where a PEV requiring charging has a number of different options of where to recharge within a reasonable driving distance. Since charging times are longer than the refueling times required by conventional vehicles, it is now of paramount importance to assign PEVs to CSs in a balanced fashion, as failure to do so may give rise to unacceptably long queues (as exemplified in Fig. \ref{fig:context}). In practice, some drivers may be interested in minimizing the driving distance (e.g., if the State of Charge (SOC) of the battery is low), some may want to minimize the charging price, or the charging time (i.e., taking into account the possible presence of queues at a CS), or in general a weighted combination of all the previous aspects. In this section we describe how a vehicle can be automatically assigned to the most convenient CS according to her/his preferences.}

\subsection{Assumptions}
\label{Assumptions}
Our work builds upon the following assumptions, that are common in most PEV-CS assignment problems:
\begin{itemize}
\item
We use a Poisson process to model a new PEV requiring charging (as in \cite{bibref:hausler} and \cite{bibref:cheki2});
\item
We assume that the time between the PEV charging request and the recommendation of the optimal CS is short enough so that the optimal solution is computed upon updated information (this implies that the assignment optimization problem is solved in a very short time, e.g., 1 s);
\item
We assume that the time required for charging is proportional to the required energy (as in \cite{bibref:hausler}, and similar to \cite{bibref:ma} and \cite{bibref:zou} with constant charge step);
\item
Finally, we assume that once a vehicle accepts to get charged to the recommended CS, it will in fact drive towards the CS. This information is used to update the total time for charging that will be required by other vehicles that intend to drive to the same CS. This assumption of compliance will then be relaxed and further analyzed in Section \ref{Compliance}.
\end{itemize}
We now describe in Section \ref{opt_prob} the proposed assignment algorithm based on personalized \textcolor{black}{cost} functions. In Section \ref{sec:det}, it is shown how it can be solved in a centralized ideal framework where all CSs are willing to exchange the relevant information, while in Section \ref{sec:stoc} we show how it can be solved in practice in a decentralized way.

\subsection{Optimization problem}\label{opt_prob}
We consider \textcolor{black}{cost} functions that are a weighted sum of three aspects: the charging time $T_{t}$, which includes the travel time to the CS, the possible time spent queuing and the effective time for charging; the charging price $P_{t}$, where we use the amount of energy generated from renewable sources as a proxy for the discount in the price of charging; and finally, the distance $D_{t}$ from the CS, used as a proxy for battery discharge (i.e., if the battery level is very low, one may just want to choose the nearest CS). In particular, such three quantities are computed and normalized for PEV $i$ as follows:
\begin{itemize}
\item $T_{t}(i) = \big(E_{t}(i) /e_r +Tr_t(i) \big)/M_{max}$ is the total time required for charging. 
This consists of the time $Tr_t(i)$ required to reach the $i$-th CS, plus the time required for charging all vehicles in the queue (included the vehicle itself), computed as $E_{t}(i) /e_r$, where $E_{t}(i)$ is the sum of all the queued energy at CS $i$ and $e_r$ is the energy received in 1 second, in kWh. The term $M_{max}$ is included for normalization purposes, so that on average this term of the \textcolor{black}{cost} function has a similar weight of the other two terms;
\item $P_{t}(i)$ represents the \emph{normalized price} of charging at CS $i$. 
In this work the charging price is computed as a function of the locally generated RE, assuming that the energy locally generated from renewable sources, if available, is free. To simplify the computation of this factor, inspired by \cite{bibref:jin}, we compute the energy price $S_t(i)$ as the following:
\begin{equation}
\label{eqn:price}
S_t(i)= e \cdot \max(m - RES_t(i), 0 ),
\end{equation}
where $e$ is the energy tariff per kWh ($[\euro/kWh]$, assumed to be the same for all CSs) for conventional energy, $m$ is the energy requested by the user, and $RES_t(i)$ is the forecast of power generation from local RE (at CS $i$) during the future charging time interval. 
Consequently, the normalized term related to the energy price in the \textcolor{black}{cost} function, will be the following term 
\[
 \textcolor{black}{P_{t}(i) = S_t(i)/(e \cdot m) = \max(m - RES_t(i), 0 ) / m. }
\]
\item $D_{t}(i) = dist(x_{CS}(i), x_{PEV}(t))/d_{max}$, where $x_{CS}(i)$ is the position of the CS $i$, $x_{PEV}(t)$ is the spatial location at step $t$ of the PEV, and $d_{max}$ is the maximum considered distance between the vehicle and a CS.
\end{itemize}
The three components are then aggregated in a single cost function as follows
\begin{equation}\label{eqn:obfun}
F_t(i) = \alpha _1 T_{t}(i) + \alpha_2 P_{t}(i) + \alpha_3 D_{t}(i),
\end{equation}
where, based on their own priorities, users can choose the values of the weights in the vector $\alpha$ ($\alpha^\top = \left[\alpha_1, \alpha_2, \alpha_3\right]$), where the three positive weights sum up to $1$.

Finally, if $I$ denotes the set of the available CSs, then at every time step $t$ when a PEV asks for charging its battery, the following optimization problem is solved: 
\begin{equation}
\label{eqn:p_det}
i^* = \argmin _{i \in I } F_t(i),
\end{equation}
where the outcome $i^*$ is the most convenient CS to be recommended to the PEV, and the \textcolor{black}{cost} function $F_t(i)$ is defined as in Equation \eqref{eqn:obfun}.

The role of the $\alpha$ parameters in Equation \eqref{eqn:obfun} is crucial because they represent the preferences of the drivers, and consequently drivers that choose different values of $\alpha$'s will in principle receive different assignment recommendations. The normalization factors are required to make the different objectives comparable, so that when one driver gives the same importance to all three the components (i.e., equal weights), the three components have a similar impact on the recommended CS. \\
\newline
\textbf{Remark: } While any other more sophisticated equation may be used to compute the charging price, Equation \eqref{eqn:price} simply reflects the fact that power generated from RE is cheaper than the power generated from other sources.

\subsection{Centralized solution}\label{sec:det}

The problem \eqref{eqn:p_det}-\eqref{eqn:obfun} is a discrete optimization problem that can be easily solved in a centralized way by checking what value of $i$ (i.e., what CS) gives rise to the lowest value of the \textcolor{black}{cost} function. PEVs looking for charging broadcast their position, personal preferences and energy requests, and the algorithm computes the optimal CS solution for them; however, this requires that different CSs exchange some personal relevant data (e.g., the length of the queue at each CS, and the expected future availability of energy generated from renewable sources at each CS). In practice, CSs may be reluctant to reveal and communicate this kind of information \textcolor{black}{as this would correspond to reveal to other competing CSs how successful they are in attracting PEV owners (from the information of the length of the queue), and their pricing strategy (e.g., in terms of how much energy they generate from renewable sources)}. Also, it computationally inefficient to continuously communicate such information. For this reason, in the next section we show how the same problem can be solved in a decentralized way, exploiting the same strategy developed in \cite{bibref:hausler}, and in Section \ref{sec:results} we shall use the solution obtained in the centralized framework as a benchmark for comparison.

\subsection{Decentralized solution} \label{sec:stoc}
A decentralized stochastic implementation of the previous algorithm usually has a number of advantages, as mentioned in \cite{bibref:hausler}. In particular, if the centralized-deterministic approaches require a large amount of communication between all participants, in the stochastic algorithm these kinds of requirements are usually lower; CSs do not need to exchange information among themselves (which may be convenient in terms of privacy preservation of some relevant data). Finally, decentralized solutions are known to be more robust than centralized solutions in general.\\
\newline
Here, we use the same approach proposed in \cite{bibref:hausler} to convert the centralized algorithm in one that can be solved in a decentralized manner. 
In practice, we assume that when a PEV needs charging, it broadcasts (to all CSs) its position, its personal preferences (i.e., vector $\alpha$ in Equation \eqref{eqn:obfun}), and its energy request. 
Then, the CSs start to broadcast a green signal (GS) with a frequency that is proportional to the value of the \textcolor{black}{cost} function. 
Mathematically, the green signal's frequency is modeled by a decreasing function of the objective function $F_t(i)$ defined in \eqref{eqn:obfun}. 
In this way, the \textcolor{black}{most convenient CSs communicate} their availability more frequently. 
In particular, we assume that the $i$-th CS communicates its availability at time step $t$ with probability $p_{CS} ^{(i)}$, that similarly to \cite{bibref:hausler} is computed as
\begin{equation}
\label{eqn:p_CS}
p_{CS} ^{(i)} (t) = 10^{-   F_t(i)},
\end{equation}
where $F_t$ is the objective function (as in Equation \eqref{eqn:obfun}), computed with the data of the PEV.
Once a vehicle receives a green signal by a CS and accepts the recommendation, then it travel towards the recommended CS, and its required energy is added to the queue of the chosen CS. 
\textcolor{black}{Thanks to the Theorem 1 of \cite{bibref:hausler}, we are also able to estimate the expected waiting time before the first green signal is received, which does not exceed 2 seconds.}
\textcolor{black}{Algorithm~\ref{Algorithm_pseudocode} gives the pseudocode of the decentralized approach just described.}

%
%

\begin{algorithm}
  \caption{\color{black} Decentralized algorithm for the PEV assignment to the CSs}
  \label{Algorithm_pseudocode}
  \begin{algorithmic}
    \color{black}
    \footnotesize
    \REQUIRE at $t$, the PEV broadcasts $\alpha$, $x_{PEV}(t)$, and the required energy $m$;
	\FOR{$\forall i \in I$}
	\STATE
		$F_t(i) = \alpha _1 T_{t}(i) + \alpha_2 P_{t}(i) + \alpha_3 D_{t}(i)$, \\
		$p_{CS} ^{(i)} (t) = 10^{-   F_t(i)}$
	\ENDFOR
    \WHILE{the PEV is not assigned}
		\FOR{$\forall i \in I$}
    			\IF{rand(1) $ < p_{CS} ^{(i)}$}
   				\STATE 
				CS $i$ broadcasts a Green Signal (GS)
			\ENDIF
		\ENDFOR
	\IF{$\exists i^*$ s.t. CS $i^*$ has emitted a GS}
	\STATE
	the PEV is assigned to CS $i^*$ (if more than a GS is received at the same time instant, the nearest one is chosen)
    \ELSE
    \STATE
    PEV waits for the GS
    \ENDIF
    \ENDWHILE
  \end{algorithmic}
\end{algorithm}

\textbf{Remark: }While both the centralized and the decentralized solution solve the same problem \eqref{eqn:p_det}-\eqref{eqn:obfun}, the centralized solution is guaranteed to recommend the most convenient CS. On the other hand, in the decentralized solution the most convenient CSs advertise their availability more often than the least convenient CSs, so it is more likely, but not guaranteed, that the PEV will first sense the availability of the most convenient CS.

{\color{black}
\section{System Architecture and Compliance}
\label{sec:DLTToken}
In the previous section we described the design of a decentralized system whose aim was to optimally assign (according to individual cost functions) electric vehicles to charging stations. 
\begin{figure*}[h!]
\centering
\includegraphics[width=2\columnwidth]{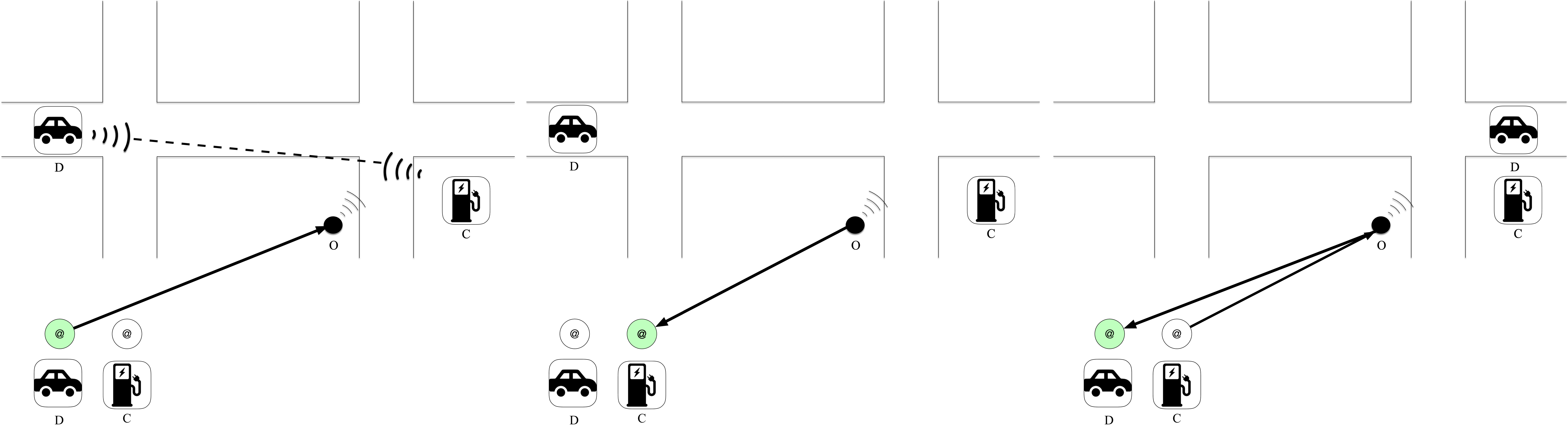}
\caption{(Left Panel) Driver D matches with station C and deposits token for initiating a contract as they are added to C's queue. (Middle Panel) Token is added, by validation of observer O to the account of station C. (Right Panel) Once the driver D reaches station C, observer O registers that the contract is fulfilled and activates the conditions, for the token to be returned to its owner.}
\label{fig:compliance}
\end{figure*}
This scheme, while appearing simple in its structure, presents one main technical challenge that needs to be addressed: \textit{How does one make sure that, after accepting a recommended assignment, users will in fact travel to the assigned CS?}. 
In this regard, it is worth noting that the issue of compliance is not a well addressed problem in studies of algorithms to regulate, control, and optimize city infrastructures \cite{oldpaper}. 
In this specific manuscript with non compliance we refer to any agent (we will use the word agent and user interchangeably in the rest of the paper) signing up for station $i$ and then going to another station (while at the same time occupying the position in the queue with station $i$). Of course, in a setting where the length of the queue affects the assignment of other vehicles, this kind of behaviour is expected to affect the performance of the proposed algorithm. Simulations are provided later in the paper, to confirm the negative effects of non compliance on the system.\newline
To address the compliance issue, we explore the use of Distributed Ledger Technology (DLT), and in particular the use of a permissioned DLT based on Directed Acyclic Graphs (DAGs). The basic idea, briefly depicted in Fig. \ref{fig:compliance} and explored more in detail later in this section, is to use token bonds, in the form of a cryptocurrency, to enforce users to comply with the matchmaking system. There are a number of reasons to use a DAG-based DLT architecture for the aforementioned task and, in general, as a communication layer in a smart city environment: \newline
\begin{itemize}

\item {\em Privacy:} In DLTs, transactions are pseudo-anonymous. This is due to the cryptographic nature of the private address\protect\footnote{https://laurencetennant.com/papers/anonymity-iota.pdf}, that allows single individuals to protect their privacy behind a large number of transactions \cite{cryptoprivacy}. This does not mean that DLTs users' identities are completely anonymous as it is possible to trace back the trail of transactions among addresses. However, through randomization of the address, it is possible to make it expensive for malicious entities to trace the transactions. This makes DLTs more robust, from a privacy perspective, in an IoT and smart mobility scenario.\newline

\item {\em Ownership:} Transactions in DLTs can be encrypted. This allows every agent to decide who can have access to the data present in their own transactions. In our setting the only data required to remain public is the ownership of the tokens, whereas other information (e.g., user quality of service, statistics on the usage of the system) can be encrypted. This allows each user to maintain ownership of their data and to use them as they please (e.g., to monetize them at a later stage).\newline

\item {\em Microtransactions:} Due to the large amount of agents in an urban environment, and due to the need of rapidly changing real time conditions (such as prices of energy, state of the energy grid, urban pollution), data need to be exchanged at a fast and large throughput. Public Blockchain architectures fall short in terms of the speed required for the CS assignment problem, whereas DAG-based ones may overcome this limitation.\newline

\item {\em Fees:}	 The compliance scheme proposed in this section relies heavily on the use of a feeless DLT (such as the IOTA Tangle \cite{Popov}). In fact usual payment systems (e.g., VISA, Mastercard,…) and classical Blockchain architectures (e.g., Bitcoin, Ethereum), due to their intrinsic design, will require users to pay a fee for every transaction performed. This, as will be explained later, makes them both inadequate to serve as the backbone for the proposed compliance scheme. \newline
\end{itemize}

Finally, the complete architecture of the system is shown in Fig. \ref{fig:architecture}. It comprises two layers: a physical one, where the operations described in the previous section take place and a compliance layer, in the form of a DAG-based DLT and a feedback controller, whose aim is to regulate the compliance of the PEVs to the desired level. In the remainder of this section we introduce and describe the structure of the DLT system and the feedback controller.\\

{\bf Remark:} Notice that the separations between the two layers is purely from a functional point of view. Due to the focus of this paper, in fact, we are limiting the current discussion on the use of DLT for the compliance one. In practice, it would be possible to use the DLT system as a more general communication system that could be employed to also carry out operations involved in the physical layer.
\begin{figure}
\centering
\includegraphics[width=\columnwidth]{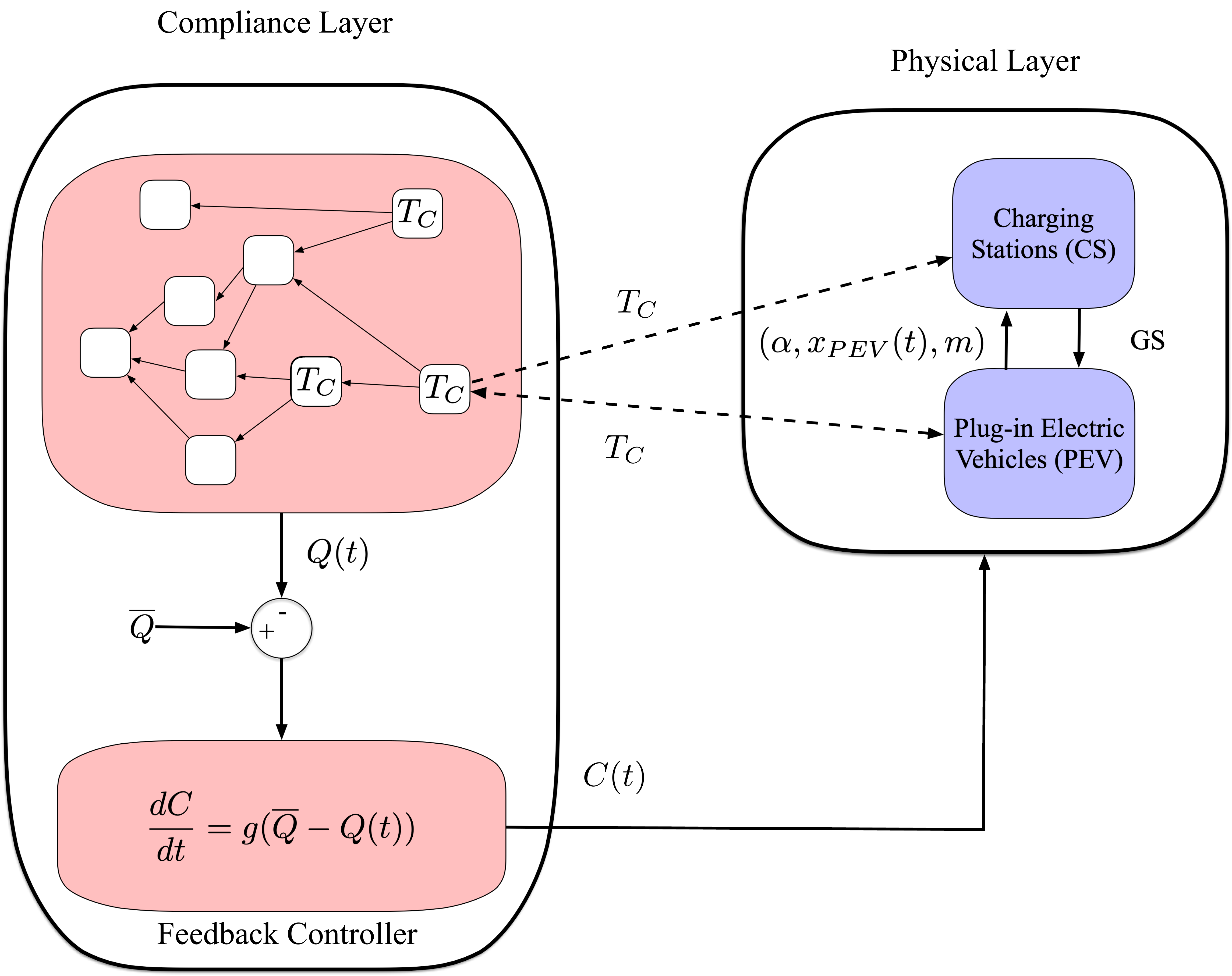}
\caption{{\color{black}Architecture of the overall system. The physical layer takes care of the matching between vehicles and CSs. The compliance layer is used to enforce compliance between users in the form of a token bond, registered through transactions on a DLT. PEVs deposit token bonds, $T_C$, on the CSs accounts and they retrieve it, if they complied with the matchmaking system (refer to the dashed lines).The value of the token bonds, $C(t)$ is regulated by a feedback controller, whose aim is to stabilise the compliance lever, $Q(t)$, around a desired level $\overline{Q}$.}}
\label{fig:architecture}
\end{figure}
{\color{black}\subsection{DAG-based DLT}
\label{Compliance}

The acronym DLT is a term that describes Blockchain and a suite of related technologies. From a broad perspective, a DLT is nothing more than a ledger held in multiple places, and a mechanism for agreeing on the contents of the ledger, namely the consensus mechanism. Made famous in the white paper from Nakamoto in 2008 \cite{Nakamoto}, the idea of Blockchain has shown great promise, especially as an immutable record keeping tool that enables financial transactions based on peer-to-peer trust \cite{Puthal}. In order to reach consensus, architectures such as Blockchain operate a probabilistic election leader mechanism enabled via a system called Proof of Work (PoW) \cite{Nakamoto}, whereas other architectures such as the IOTA Tangle are based on DAG structures and a swarm like strategy to achieve consensus \cite{Popov}.  A DAG is a 
finite connected directed graph with no cycles. In other words, in a DAG there is no directed 
path that connects a vertex with itself. The IOTA Tangle is a particular instance of a DAG-based DLT 
\cite{IoTPaper}, where each transaction is represented by a vertex, and where the 
 topology of the graph represents the ledger. Refer to Fig. 
\ref{Fig: Tangle} for a brief description of this process and to  \cite{oldpaper} for a more thorough analysis.
As anticipated before, the Tangle has the advantage over Blockchain to allow microtransactions without any fees 
(as miners are not needed to reach consensus over the network \cite{Popov}).}

\begin{figure}
    \centering
    \includegraphics[width=0.7\columnwidth]{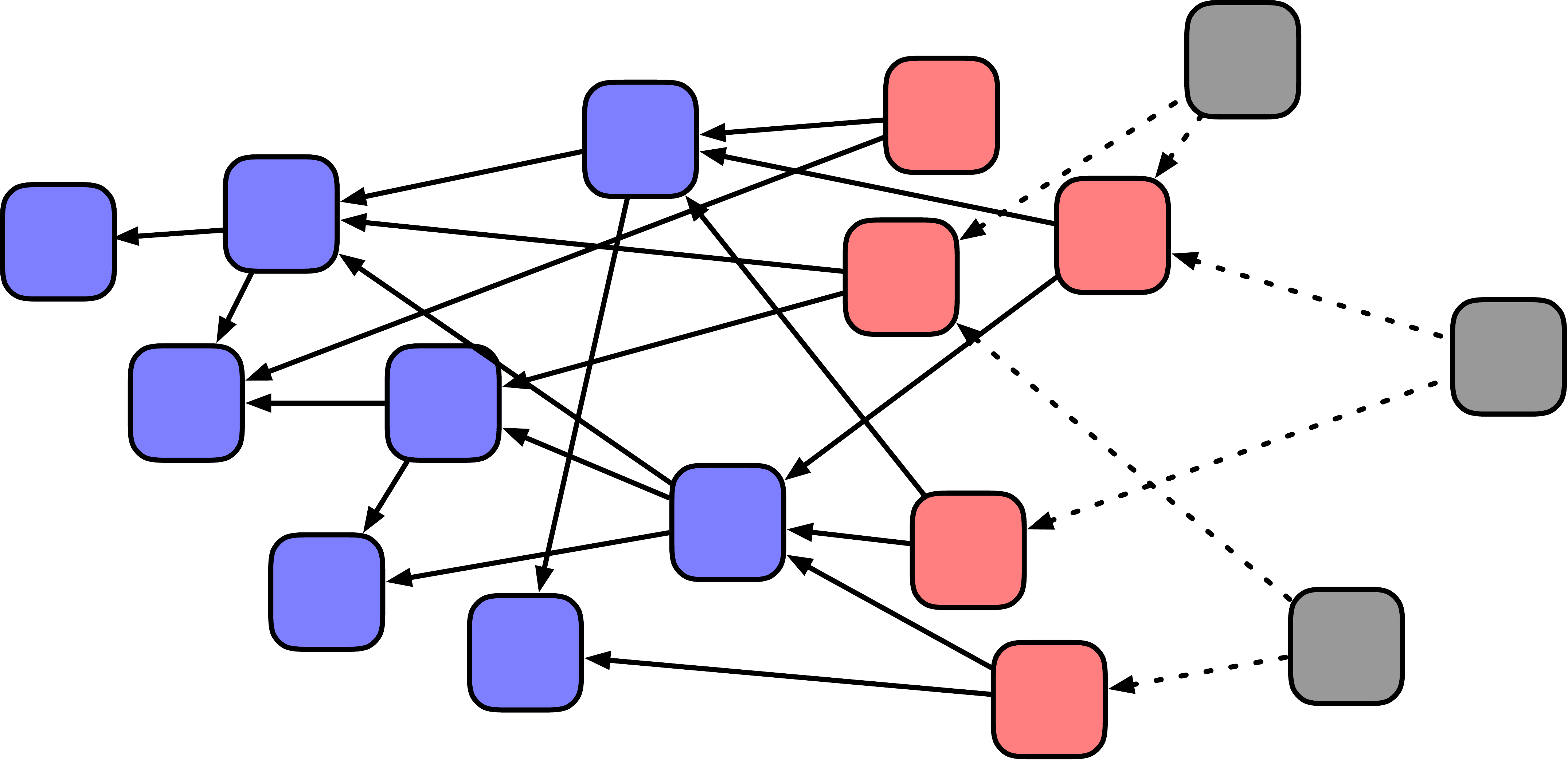}
    \hspace{10mm}
    \includegraphics[width=0.7\columnwidth]{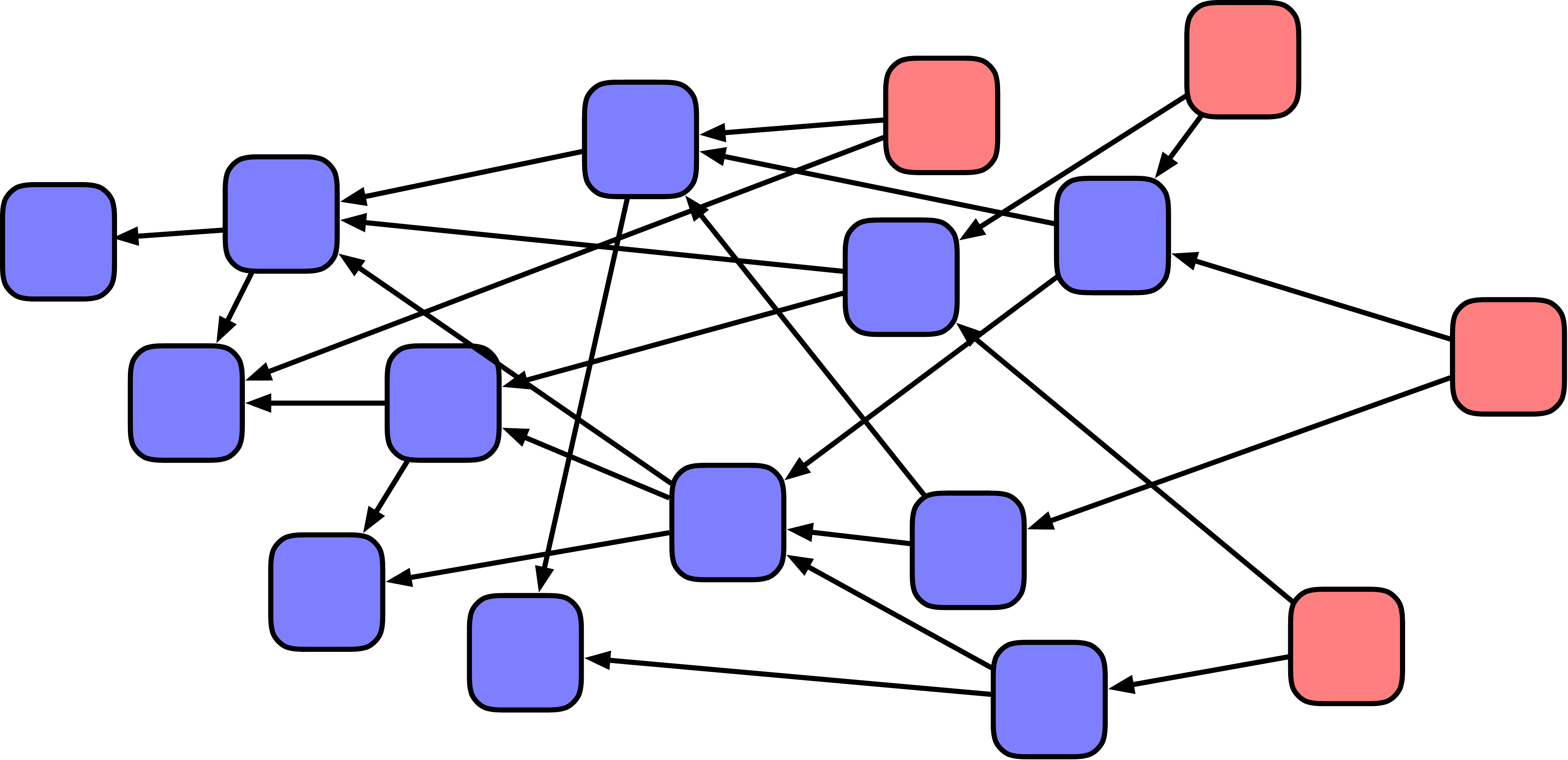}
    \caption{{\color{black}Sequence to issue a new transaction in the Tangle. In the top panel, 3 new transactions (grey vertices) are issued and they are in the process of carrying out their PoW. In the bottom panel the PoW has been completed and those vertices are added to the Tangle. The blue sites represent the approved 
    transactions while the red ones describe transactions that have not received any approval yet. 
     Black full edges represent approvals, whereas dashed ones represent transactions that 
    are in the process of completing the PoW.}}
    \label{Fig: Tangle}
\end{figure}

\subsection{Compliance Mechanism}

The concept behind compliance was already explored in \cite{oldpaper} and revolves around the use of \emph{crypto tokens}, associated with some economic values, to be used as collateral in order to be assigned to a specific queue from a CS. Refer to Fig. \ref{fig:compliance} for a visual representation of this process. Whenever there is an agreed match between a vehicle and a station, the vehicle deposits some amount of tokens into the charging station account and, when the vehicle shows up to the station, the tokens are automatically returned. Notice that if this operation was performed on a classical Blockchain or on other common payment methods, the user would be forced to pay a fee for each transaction and, due their nature as a bond, which needs to be returned in its full value to the user (given that the user complied), a fee on each transaction would make this system non feasible.\newline
From a formal point of view this process can be described as follows: we are interested of maintaining the overall level of compliance of the system to a desired level $\overline{Q} \in [0,1]$. To do so, we introduce a controller that regulates the value of the deposit bond, $C(t)$, using the current compliance level $Q(t)$ (this value is extracted directly from the ledger). Refer again to Figure \ref{fig:architecture} for a better understanding of the interconnection between each layer. \newline

Finally, the compliance layer is designed in such a way that whenever a user issues a token, this bond can be retrieved by the user \emph{if and only if} the vehicle reaches the CS in a certain amount of time. This process can be made automatic by using a combination of smart contracts (a computer program that executes as soon as certain conditions are met) \cite{ethereum} and an access mechanism called Proof of Position (explained later in the text) \cite{SPtoken}: namely a DAG-based DLT called \emph{Spatial Positioning Token} (SPToken) which represents a permissioned version of the IOTA Tangle. This means that, unlike public DLTs, in which each user has complete freedom on how to update the ledger with transactions, the SPToken network has a regulatory policy based on the physical positions of agents. This feature allows for a number of different uses: it can be employed  to prevent agents from spamming transactions that do not possess any relevant data (since transactions can be encrypted)\cite{SPtoken} or, as in this specific application, it can be used to make sure that an agent satisfies certain conditions and therefore that cheating is not possible from both sides of the transaction (e.g., a dishonest CS might be inclined to not return the bond even though a user complied with the matchmaking system). Therefore, as a validation mechanism, SPToken makes use of, beside the classical PoW, also of a PoP to authenticate transactions. In other words, for a transaction to be authenticated, it has to carry proof that the agent was at a certain location. This is achieved via special nodes called \emph{Observers} (see Fig.~\ref{fig:compliance}). For a thorough discussion on this topic, the interested reader can refer to \cite{SPtoken}.

}

\section{Analysis and Results} \label{sec:results}
We now analyze and present the main results that are obtained by applying the proposed algorithms to assign PEVs to CSs. 
To better clarify how the algorithm works, we first present results that are obtained in an idealized scenario where CSs are points in a unit square and vehicles can directly drive within the square towards the CSs. All distances are thus measured using the Euclidean metrics, disregarding of whether there exists a direct road that connects the PEV with a CS. In the second set of simulations, we relax this assumption, and distances are the true distances in a realistic urban network.

In our simulations, we assume that vehicles are charged with a \textcolor{black}{charge rate of $22$ kW (i.e., $e_r = 0.0061$ kWh in the definition of $T_t$ in \eqref{eqn:obfun}) at each simulation step (1 sec), that is consistent with the power rate of most public charging poles  (see \cite{bibref:charge_data}). In \cite{bibref:charge_behav} it had been shown that the average charged energy at urban CSs is around $5$ kWh, and consistently we have assumed that the energy required by PEVs is sampled from a Gaussian distribution centered in $5$ kWh with a standard deviation of $1.2$ kWh}. This implies that the maximum charging time is about $1300$s (i.e., about $22$ min), corresponding to a request of $8$ kWh. Finally, we assume that CSs may either be equipped with PV panels of $10$ kW of nominal power, or with a wind turbine of $5$ kW, or may not have renewable sources at their disposal. 
\textcolor{black}{For our simulations, we have used MATLAB (2016b release) to simulate the whole system and the algorithm, with Windows 10 as operating system. The mobility simulator SUMO (v. 1.3.1) was used to simulate vehicular flows in the second set of simulations (Sections \ref{sec:sumo_res1} and \ref{sec:sumo_res2}), for the case study of the city center of Pisa, Italy.}

\textcolor{black}{
\subsection{The Effectiveness of the Multi-Objective \textcolor{black}{Cost} Functions}\label{sec:kpi}}
In our paper we consider personalized \textcolor{black}{cost} functions. This implies that two drivers with different priorities may have two different recommended  CSs.

\begin{figure}[h!]
\centering
\includegraphics[width=1\columnwidth]{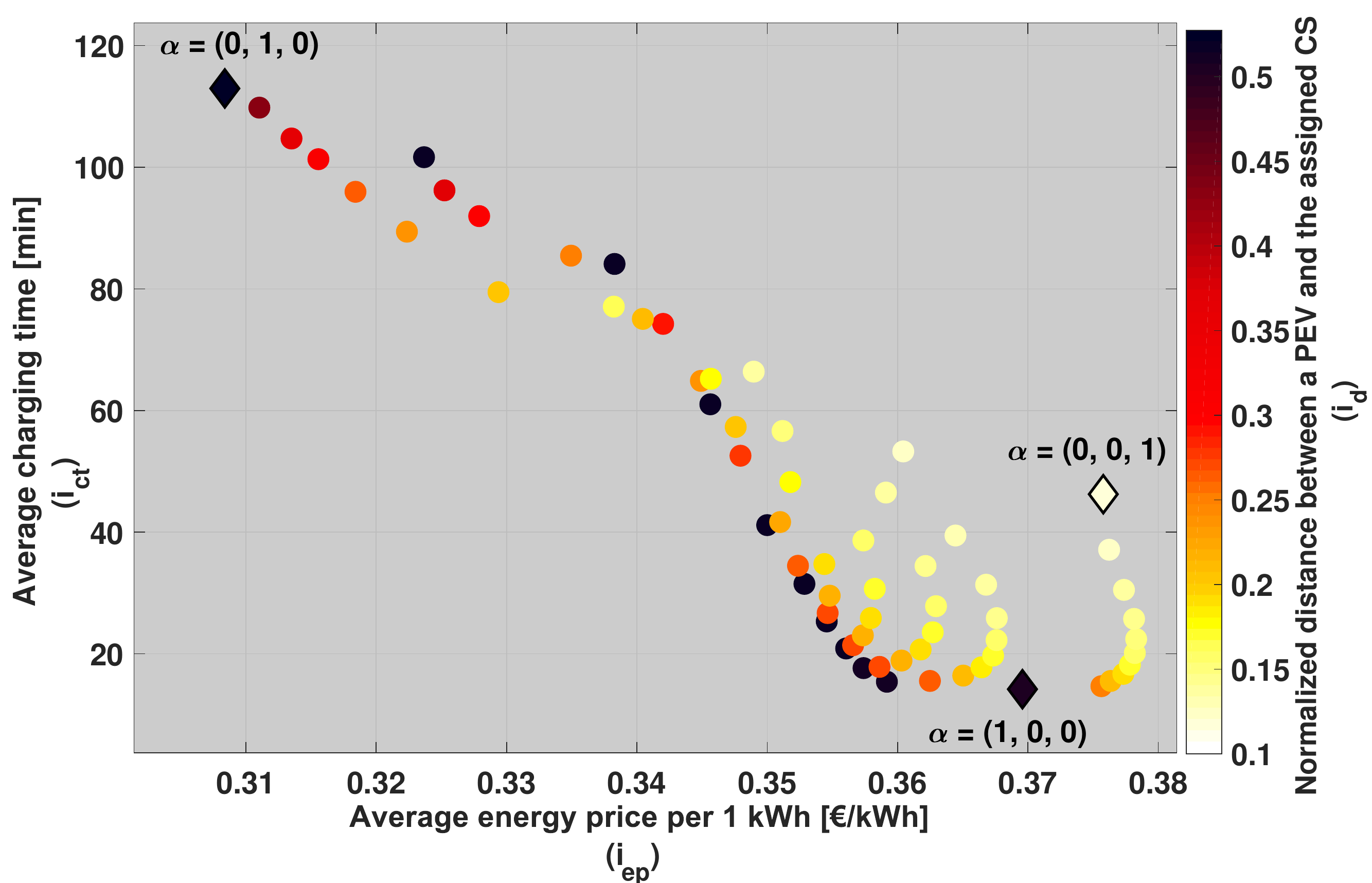}
\caption{Plot of the performance indices for $66$ different combinations of the convex 3-tuple $(\alpha_1, \alpha_2, \alpha_3)$, with a step size of 0.1. The diamonds correspond to extreme cases when only one of the three drivers interests considered is minimized.}\label{fig:all_KPI}
\end{figure}

In order to clarify the exposition, we now introduce three different performance indicators to quantify the impact of the single components of the \textcolor{black}{cost} function \eqref{eqn:obfun}:
\begin{enumerate}
\item $i_{ct}$, the performance index for the charging time. It is the average waiting time for charging (i.e., it includes travel time to the CS, the possible time spent queuing and the effective time for charging), averaged over the set of charged vehicles during the simulations;
\item $i_{ep}$, the average energy price per kWh. 
It is computed following the idea of Equation \eqref{eqn:price}, that is by considering $e \cdot \Big(E_{charged} - E_{RE} \Big)/E_{charged}$, where 
the tariff $e$ is chosen as $0.45$ $\euro/kWh$, which is consistent with the price in Italy\footnotemark; 
$E_{charged}$ is the energy charged in the simulation, and $E_{RE}$ is the amount of RE used.
\footnotetext{https://www.enelx.com/it/it/mobilita-elettrica/prodotti/privati/servizi-x-recharge}
Consequently, in the absence of energy generated from REs, the average energy price per kWh will be exactly $0.45$ $\euro/kWh$, whereas if the CSs have their own REs, they may sell energy at a lower price. 
\item $i_{d}$, the average distance between a PEV and the assigned optimal CS (here, in the unit square).
\end{enumerate}
For the next analysis, we considered results related to \textcolor{black}{75 Monte Carlo simulations run with different cars positions and different charging requests}; the performance indicators are computed and averaged over the 75 different simulations, in order to get a more robust statistical index. Each simulation is related to a period of 7 hours, during which about 500 PEVs get charged. \\
Assuming that all the drivers have exactly the same weighting coefficients in Equation \eqref{eqn:obfun}, Fig. \ref{fig:all_KPI} illustrates the 3D plot when the weights vary between 0 and 1, with a step size of 0.1. 
Since only combinations of weights that sum up to 1 are considered, then 66 overall cases appear in the figure. 
In particular, diamonds are used to show the extreme cases when all users decide to optimize a single component of the \textcolor{black}{cost} function. 
Among other things, it is possible to observe that charging times rise to about 110 minutes when all vehicles aim at minimizing charging costs as all PEVs go to the cheapest CS, while charging times reduce to about 16 minutes when all vehicles minimize the time required for charging. 
In this case, there are practically no queues at CSs and a very short travel time to the station, since the average time for only charging is about 13.5 minutes. 
This result further emphasizes the \textcolor{black}{importance} of smart charging, intended as a smart assignment of PEVs to CSs.

\subsection{Centralized versus Decentralized}
This second section of results compares our stochastic decentralized algorithm with the deterministic centralized one. 
Recall that, in the stochastic approach it is probable, but not guaranteed, that a vehicle is assigned to the optimal CS. 
Thus, in this simulation we wish to identify how suboptimal is the stochastic decentralized solution with respect to the deterministic centralized solution, that represents the benchmark. 

In order to simplify the interpretation of this analysis, we compare the two methodologies in one setting, which is the one referring to the parameters configuration $\alpha = (1, 0, 0)$; similar results are obtained also for the remaining 3-\textcolor{black}{tuple} of $\alpha$'s.

\begin{figure}
\centering
\includegraphics[width=1\columnwidth]{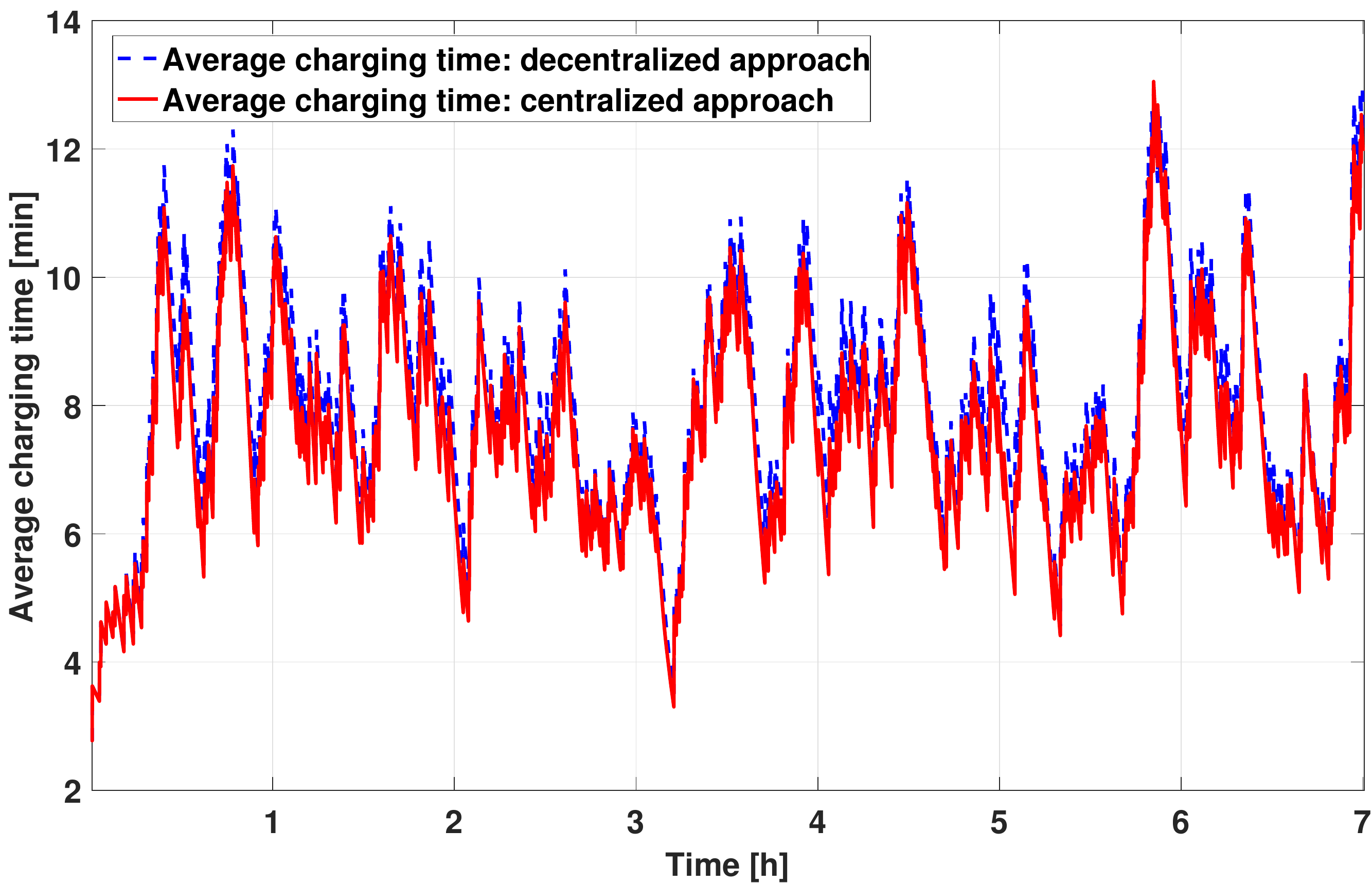}
\caption{The blue dashed curve shows the average charging time in the decentralized stochastic approach, while the red solid line corresponds to the centralized deterministic case (optimal full-information solution). 
The average is computed over all CSs, and some CSs could be empty during the simulation time.}
\label{fig:queue_100}
\end{figure}

\emph{Configuration: $\alpha = (1, 0, 0)$}: 
in this case the only goal is to minimize the charging time. Similar results to the ones of paper \cite{bibref:hausler} are obtained, since we just add the travel time to the waiting time factor of the previous model. 
\textcolor{black}{Fig. \ref{fig:queue_100} shows the charging time of the whole system as a function of the time, and it is a mean value, including possibly empty CSs}.
As can be seen from the graph, the behaviour of the decentralized stochastic approach (dashed line) is very close to the optimal one (solid line): the RMSE (Root Mean Squared error) is about $0.5087$ minutes, 
while on average the charging time of the decentralized solution is about $5$\% more of that of the centralized case ($8$ minutes instead of $7.6$ minutes). 

\begin{figure}
\centering
\includegraphics[width=1\columnwidth]{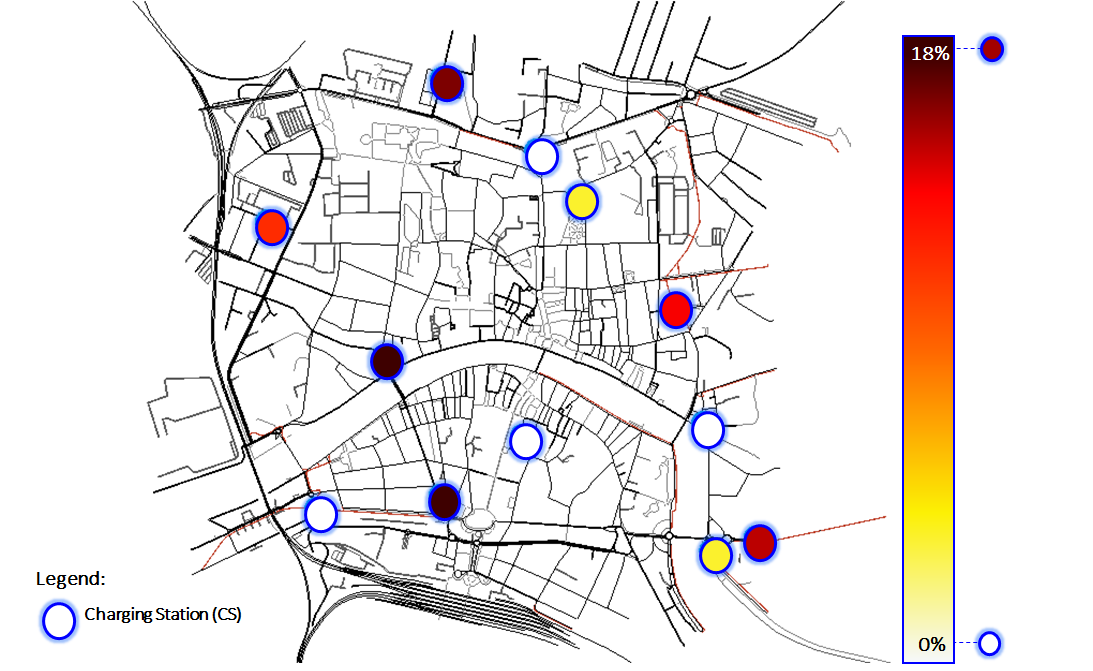}
\caption{Heat map of the city of Pisa, scenario 1: drivers interest is to minimize the charging cost. Blue circles represent the position of CSs in the city, and are colored based on their participation factor.}
\label{fig:pisa_010}
\end{figure}

\begin{figure}
\centering
\includegraphics[width=1\columnwidth]{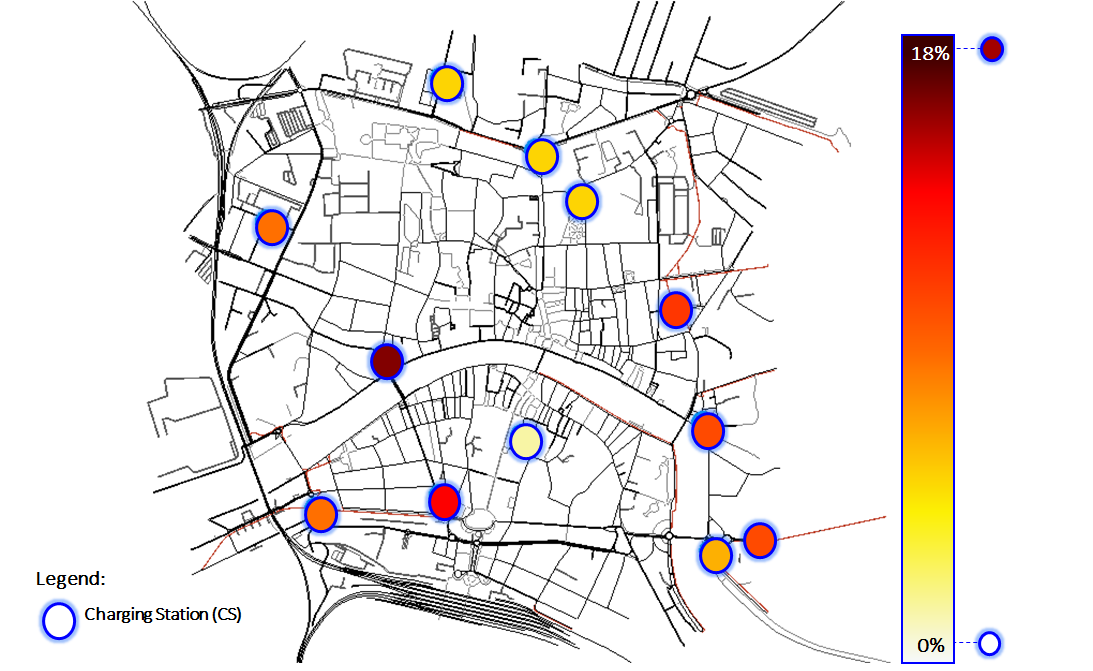}
\caption{Heat map of the city of Pisa, scenario 2: drivers interest is to minimize all the three factors, that is the charging time, the price and the distance. Blue circles represent the position of CSs in the city, and are colored based on their participation factor.}
\label{fig:pisa_111333}
\end{figure}

\textcolor{black}{
\subsection{Realistic Simulations with SUMO}\label{sec:sumo_res1}}
In this section we investigate the proposed approach using the mobility simulator SUMO in a realistic urban network. 
For this purpose, \textcolor{black}{we consider the road network of the city-center of Pisa, with 12 CSs, placed as shown in Fig. \ref{fig:pisa_010}-\ref{fig:pisa_111333}, according to the existing CSs network\footnote{\textcolor{black}{https://www.colonnineelettriche.it/}}}. 
In \textcolor{black}{the} first simulation, we mimic what could happen if our proposed assignment procedure is not used. In this case, it is realistic to assume that drivers quickly learn average electricity prices, and decide to get charged to a cheap CS along (or near) their typical driving patterns, as currently occurs for vehicles with an internal combustion engine. 
In the second simulation, our proposed assignment procedure is fully used, assuming that all drivers have the same weights for all components of their \textcolor{black}{cost} function (a similar result would be obtained if the driving population were divided into equal groups of people aiming at minimizing charging times, or charging distances, or charging prices). In this case, driving quantities such us the travel time and the driving distance are computed by using available commands in SUMO, while in practical applications of our algorithm, they can be easily recovered by using popular tools and apps (e.g., Google maps, Azure maps, etc..). \\

Final results are summarized in the two heat maps (Fig.~\ref{fig:pisa_010} and Fig.~\ref{fig:pisa_111333}). 
It can be clearly seen that in the first case most vehicles decide to get charged at the cheapest CSs, while more uniform utilization factors are achieved when the proposed procedure is used. In the second case, it happens that cheaper (or nearer) CSs are initially taken, but when queues start forming then more expensive CSs may become more attractive as queues are shorter.

\begin{figure}
\centering
\includegraphics[width=1\columnwidth]{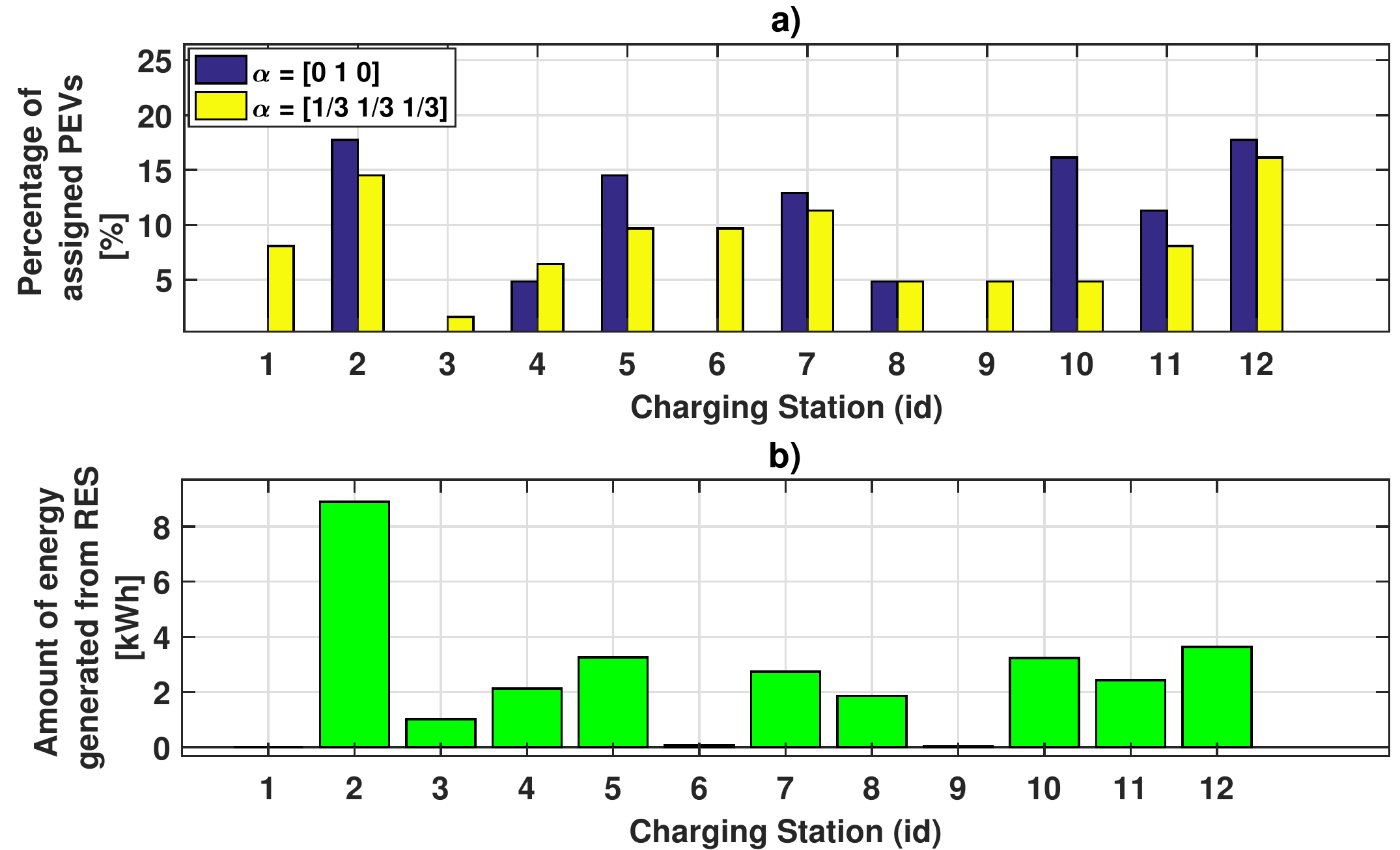}
\caption{The bar plot a) shows the participation factor, which represents the percentage of PEVs assigned to it during the simulation; in particular, PEVs interested in minimizing the costs are blue, while yellow indicates the PEVs which are interested in all components of their \textcolor{black}{cost} function. 
Bar plot b) presents the energy generated by local RES of each CS.}\label{fig:barplot}
\end{figure}

The same result can be also visualized through the bar plots shown in Fig.~\ref{fig:barplot}. In particular, it can be noticed that in the first case (drivers minimizing charging prices) all drivers choose the CSs where the prices are lower (because a greater amount of energy is generated from renewable sources, as shown in Fig.~\ref{fig:barplot}.b). On the other hand, in the second case drivers take advantage of the queuing information and by using the proposed algorithm a balancing effect is achieved among the CSs.

\subsection{Compliance Analysis}\label{sec:sumo_res2}
\begin{figure}
\centering
\includegraphics[width=1\columnwidth]{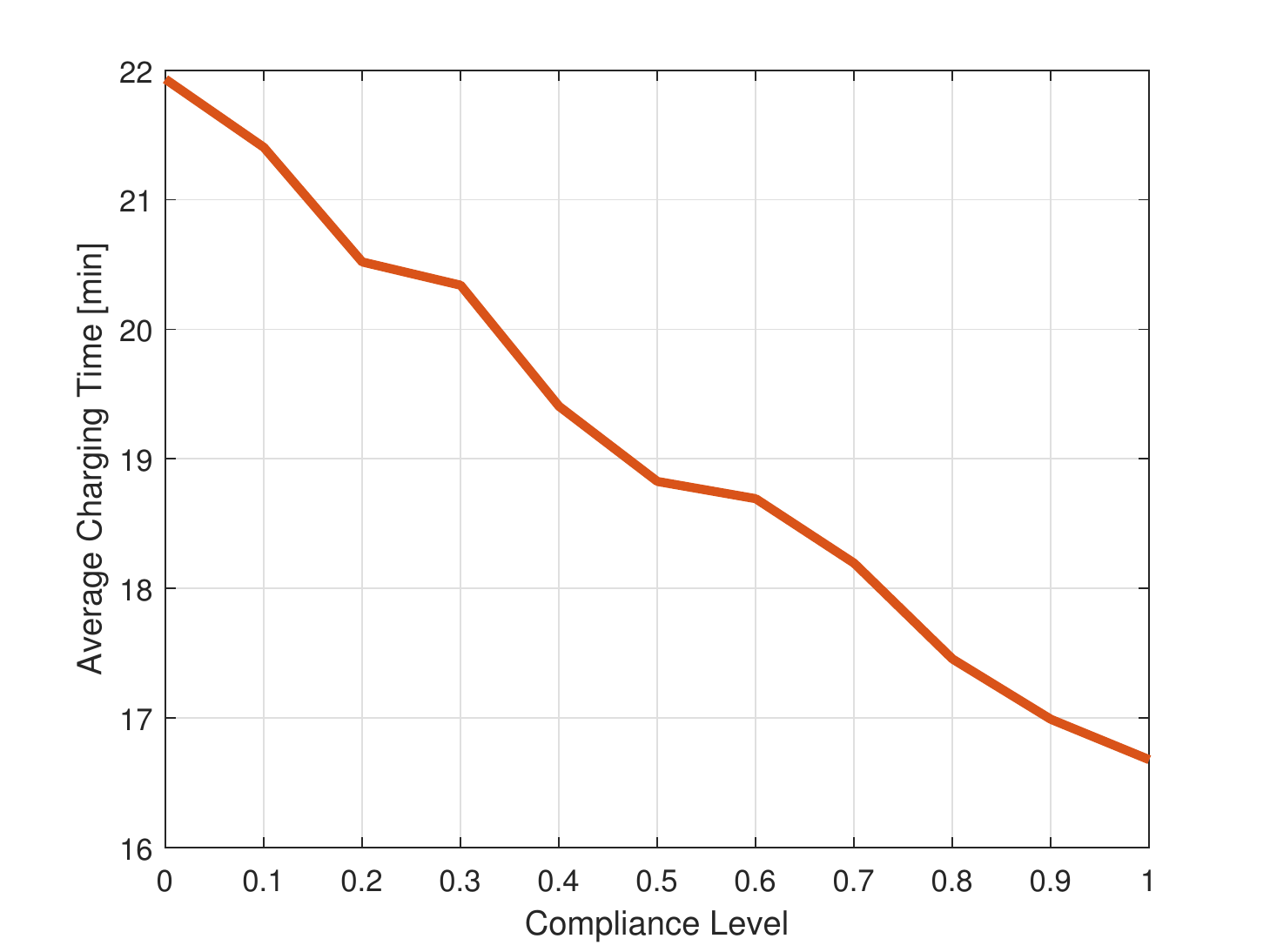}
\caption{Average charging time for a car for different levels of compliance. Results are obtained through 50 Monte Carlo simulations.}
\label{fig:ComplianceWaitTime}
\end{figure}

As a final scenario to conclude our analysis, we evaluate the performance of our algorithm in the event that a fraction $Q$ of the population is not compliant with the proposed scheme, in order to provide a rationale for the DLT-based architecture presented in Section \ref{sec:DLTToken}. To take the compliance level of the population into account, we consider the following: whenever an agent ``accepts'' the assignment to station $j$ recommended by the algorithm, the agent will in fact either comply with the accepted recommendation with probability $Q$, or it will just go to the nearest station with probability $1-Q$. Obviously, the latter event causes an inconvenience to the proposed algorithm, as the agent will still show up in the queue of station $j$. To evaluate how the parameter $Q$ affects the performance of the algorithm, we consider a scenario in which every agent tries to minimize its charging time (i.e., $\alpha_1 = 1, \alpha_2=\alpha_3 = 0$) and we show how this variable is affected when the parameter $Q$ ranges in the interval $[0,1]$. Furthermore, due to the stochastic nature of this setting, we take into account 50 Monte Carlo simulations to average out the random fluctuations and obtain meaningful values. Results are shown in Fig. \ref{fig:ComplianceWaitTime}: as it is clear from visual inspection, the average charging time decreases monotonically as the compliance level $Q$ increases. This result highlights that compliance levels and their regulation (for instance, through the use of the pricing tokens mechanism discussed in this paper) represent a critical element that, if left unchecked, might lead to low levels of quality of service in P2P applications.

\section{Conclusion and Future Lines of Research}\label{conclusion}
{\color{black}This paper described a novel procedure to assign PEVs to the most convenient CS in a fully decentralized manner. Assignments are personalized, as they take into account specific, possibly different, interests and priorities of different driver. Moreover we explored the use of an IoT architecture based on a permissioned DLT to enforce users compliance to the assignment scheme and thus to achieve a satisfactory quality of service. The proposed procedure has been illustrated through extensive simulations, and validated using the mobility simulator SUMO for the realistic case study of the city of Pisa in Italy. Simulations show the effectiveness of our approach and indicate how users compliance represents a key factor that needs to be taken into account to avoid performance degradation.}\\
\newline
The next step of this work foresees the practical implementation of the proposed algorithm. \textcolor{black}{As a first step, we shall implement the permissioned DLT, a centralized version of the algorithm, exploiting publicly available data (e.g., position and price) and the compliance mechanism. Moreover we plan to implement our system in a smartphone app that can be used by PEV drivers}. However, our plan is to involve CSs to advertise themselves their availability using private information as well (e.g., queue lengths).

\begin{IEEEbiography}[{\includegraphics[width=1in,height=1.25in,clip,keepaspectratio]{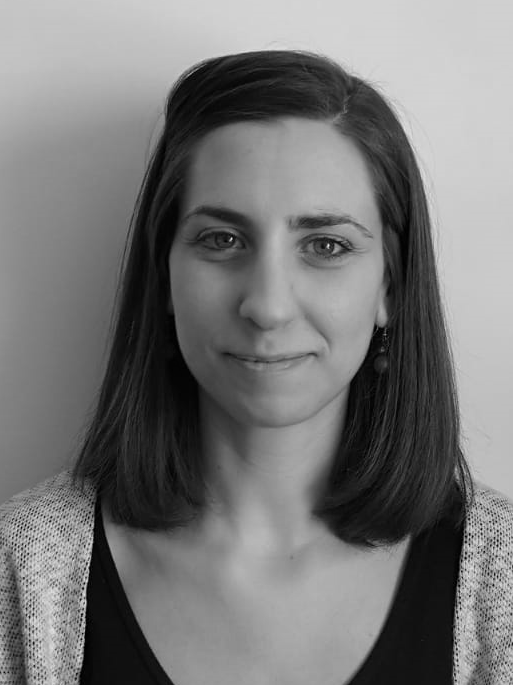}}]{Michela Moschella} (S'19) received the Master’s degree in mathematics from the University of Pisa, Italy, in 2015. She is currently a Ph.D. student with the Department of Energy, Systems, Territory and Constructions Engineering, University of Pisa. Her research interests include machine learning techniques for renewable power generation forecasting model, and optimization of large-scale systems with applications to electric mobility and smart grids.
\end{IEEEbiography}
\begin{IEEEbiography}[{\includegraphics[width=1in,height=1.25in,clip,keepaspectratio]{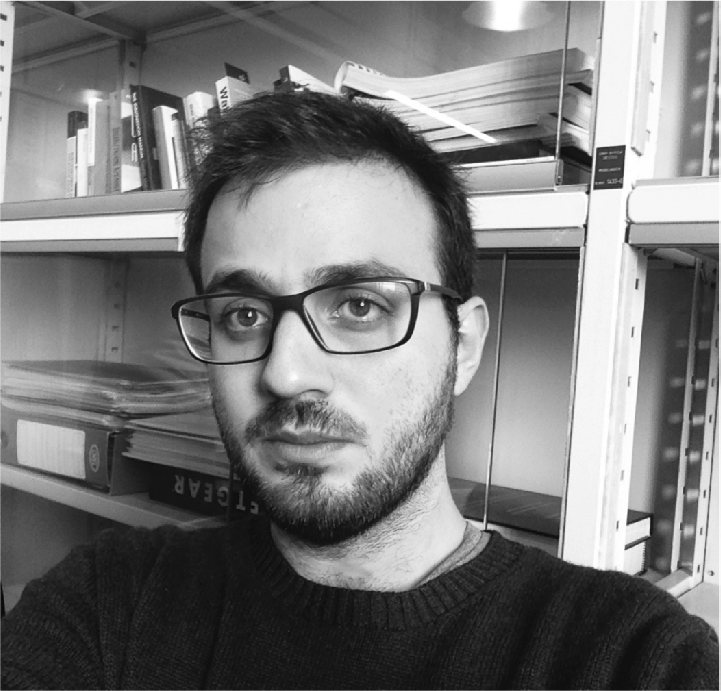}}]{Pietro Ferraro} received  the Master's degree in Robotics and control engineering and the PhD in control and electrical engineering from the University of Pisa, Italy, in 2015 and in 2018. He is currently a Research Associate with the Dyson school of design engineering at Imperial College London. His research interests include control theory applied to the sharing economy domain.
\end{IEEEbiography}

\begin{IEEEbiography}[{\includegraphics[width=1in,height=1.25in,clip,keepaspectratio]{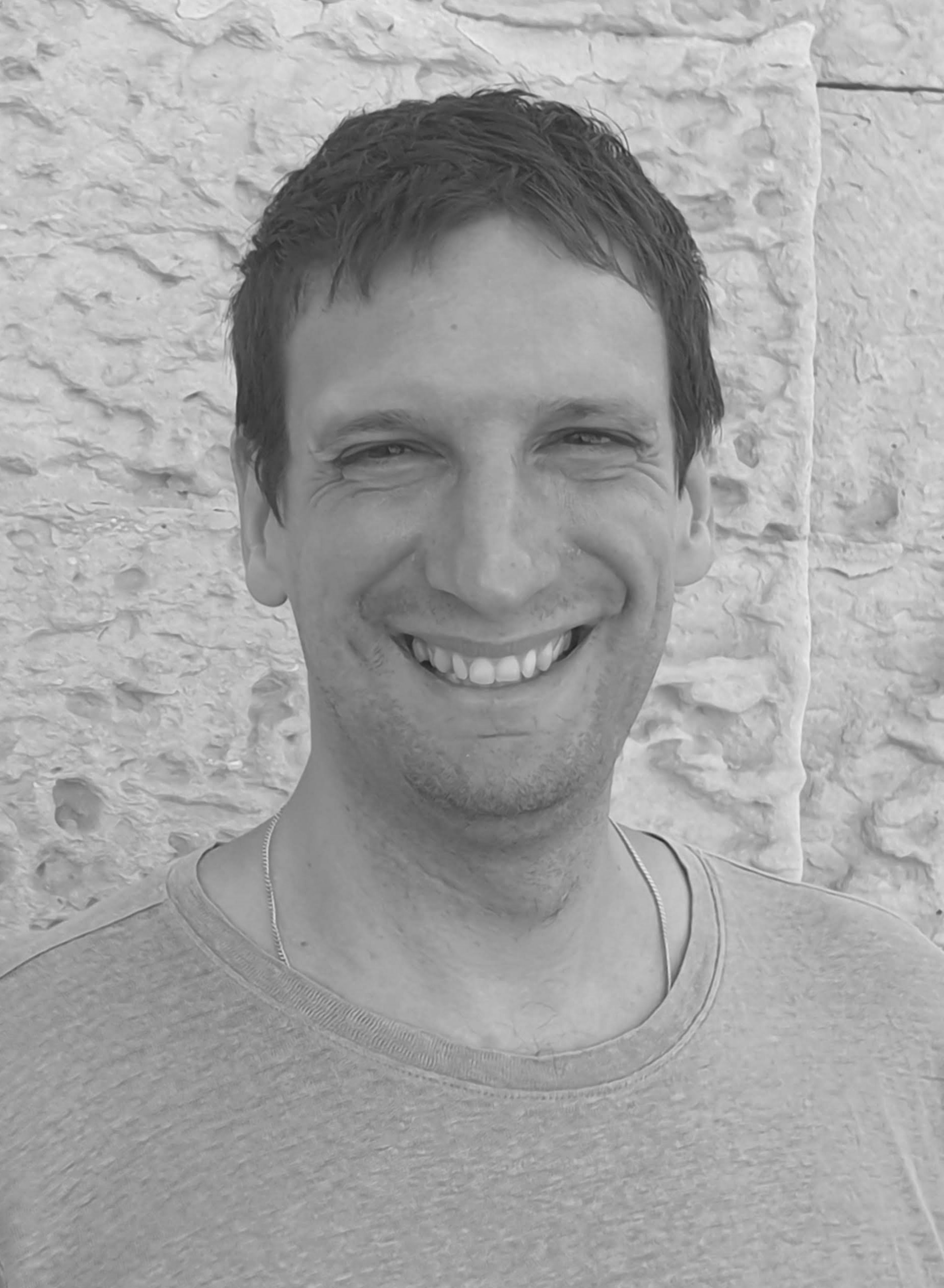}}]{Emanuele Crisostomi} (M'11--SM'16) received the Bachelor’s degree in computer science engineering, the Master’s degree in automatic control, and the Ph.D. degree in automatics, robotics, and bioengineering, from the University of Pisa, Italy, in 2002, 2005, and 2009, respectively. He is currently an Associate Professor of Electrotechnics with the Department of Energy, Systems, Territory and Constructions Engineering, University of Pisa. His research interests include control and optimization of large-scale systems, with applications to smart grids and green mobility networks. He is a co-author of the recently published book: ``Electric and Plug-in Hybrid Vehicle Networks: Optimization and Control” (Crisostomi, Shorten, Stüdli, Wirth), CRC Press, Series: Automation and Control Engineering, 2017'' and a co-editor of the upcoming Springer Nature book on ``Analytics for the Sharing Economy: Mathematics, Engineering and Business Perspectives''. 
\end{IEEEbiography}

\begin{IEEEbiography}[{\includegraphics[width=1in,height=1.25in,clip,keepaspectratio]{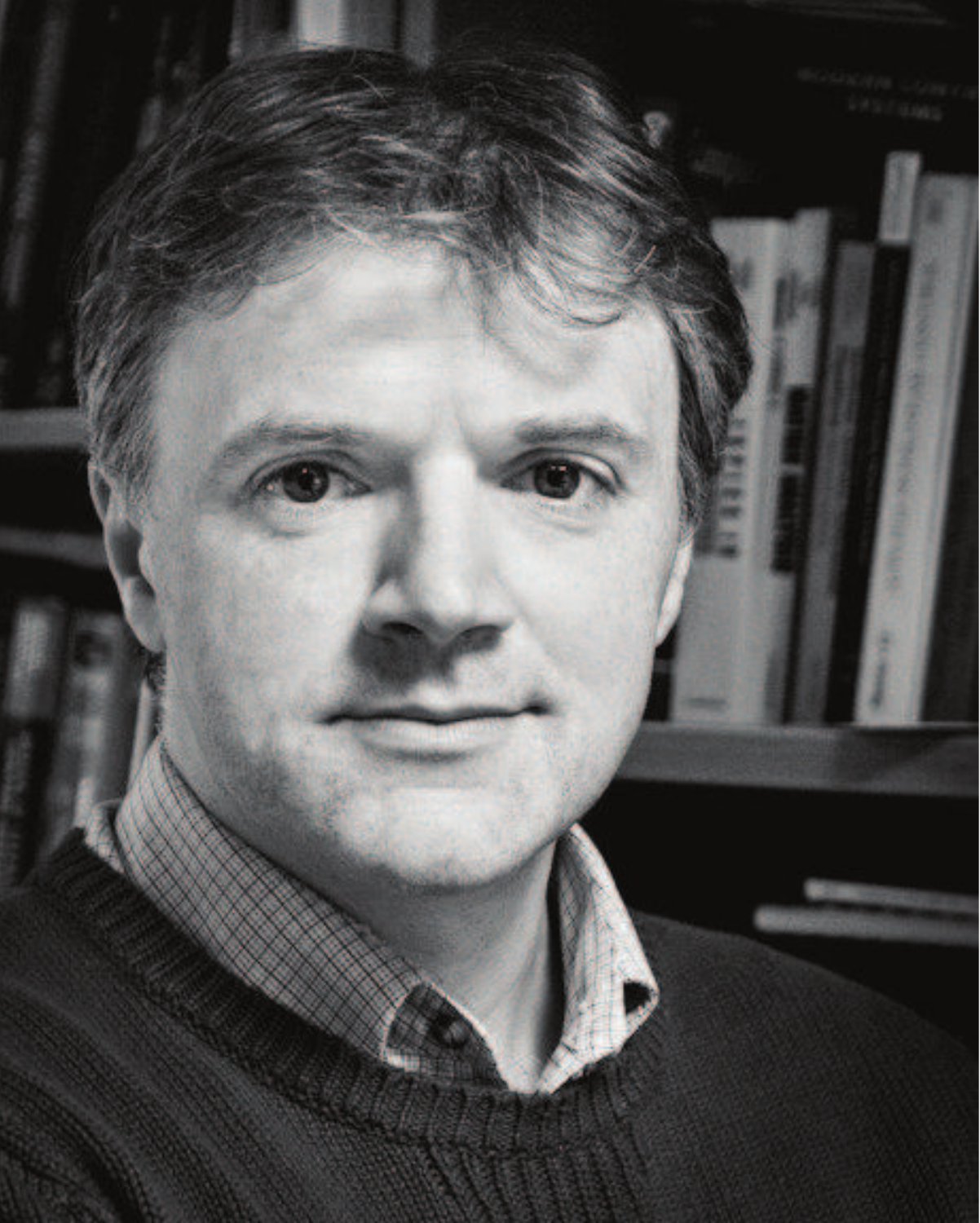}}]{Robert Shorten} graduated from UCD, with a B.E. degree in Electronic Engineering in 1990, and a Ph.D. degree in 1996. From 1993 to 1996, Professor Shorten worked at Daimler-Benz research labs in Berlin where completed his Ph.D. work, and was also the holder of a Marie Curie Fellowship. In 1996 he was invited to work as a visiting fellow at the Center for Systems Science, Yale University, commencing a long-standing research collaboration with Professor K. S. Narendra on the study of switched systems. Since returning to Ireland in 1997 as the recipient of a European Presidential Fellowship,  Professor Shorten has been active in a number of theoretical and applied research areas including: computer networking; classical automotive research; collaborative mobility (including smart transportation and electric vehicles); as well as basic control theory and linear algebra. Professor Shorten is a co-founder of the Hamilton Institute, National University of Ireland, Maynooth, where he was a Full Professor until March 2013, and was also the holder of a Visiting Professorship at TU Berlin in 2011-12. From 2013 to 2015 he led the Control and Optimization activities at IBM Research Ireland in the area of Smart Cities. He was Professor of Control Engineering and Decision Science at University College Dublin (UCD) from 2015-2019 as well as consultant to IBM Research Ireland. He is currently Professor of Cyber-physical Design at Imperial College London. Professor Shorten is a co-author of the recently published books: AIMD dynamics and distributed resource allocation (Corless, King, Shorten, Wirth, SIAM 2016). Plug-in Hybrid and Electric Vehicle Networks: Optimization and Control (Crisostomi, Shorten, St\"{u}dli, Wirth, CRC Press, November 2017).
\end{IEEEbiography}

\end{document}